\documentclass[aps,prl,twocolumn,superscriptaddress,calc,epsfig,10pt]{revtex4-1}
\usepackage{graphicx}
\usepackage{graphics}
\usepackage{amssymb,amsmath}
\usepackage{subcaption}

\usepackage{wasysym}

\usepackage{color}
\usepackage[normalem]{ulem}

\newcommand{\mitCUAaddress}{Department of Physics, MIT-Harvard Center for Ultracold Atoms, and Research Laboratory of Electronics, MIT, Cambridge, Massachusetts 02139, USA}
\newcommand{\pennstateaddress}{Department of Physics, The Pennsylvania State University, University Park, PA 16802, USA}
\newcommand{\UFRJaddress}{Instituto de Fisica, Universidade Federal do Rio de Janeiro, Cx.P. 68.528, 21941-972, Rio de Janeiro, RJ, Brazil}
\newcommand{\SanJoseAddress}{Department of Physics and Astronomy, San Jos\'{e} State University, San Jos\'{e}, CA 95192, USA}
\newcommand{\OSUaddress}{Department of Physics, The Ohio State University, Columbus, OH 43210, USA}

\newcommand{\ket}{\right\rangle}

\newcommand{\bra}{\left\langle}

\begin{document}

\title{Observation of Spatial Charge and Spin Correlations in the 2D Fermi-Hubbard Model}

\author{Lawrence W. Cheuk$^\dagger$}
\affiliation{\mitCUAaddress}

\author{Matthew A. Nichols$^\dagger$}
\affiliation{\mitCUAaddress}

\author{Katherine R. Lawrence}
\affiliation{\mitCUAaddress}

\author{Melih Okan}
\affiliation{\mitCUAaddress}

\author{Hao Zhang}
\affiliation{\mitCUAaddress}

\author{Ehsan Khatami}
\affiliation{\SanJoseAddress}

\author{Nandini Trivedi}
\affiliation{\OSUaddress}

\author{Thereza Paiva}
\affiliation{\UFRJaddress}

\author{Marcos Rigol}
\affiliation{\pennstateaddress}

\author{Martin W. Zwierlein}
\affiliation{\mitCUAaddress}
\date{\today}

\begin{abstract}
Strong electron correlations lie at the origin of transformative phenomena such as colossal magneto-resistance and high-temperature superconductivity. Already near room temperature, doped copper oxide materials display remarkable features such as a pseudo-gap and a ``strange metal'' phase with unusual transport properties. The essence of this physics is believed to be captured by the Fermi-Hubbard model of repulsively interacting, itinerant fermions on a lattice. Here we report on the site-resolved observation of charge and spin correlations in the two-dimensional (2D) Fermi-Hubbard model realized with ultracold atoms.
Antiferromagnetic spin correlations are maximal at half-filling and weaken monotonically upon doping. Correlations between singly charged sites are negative at large doping, revealing the Pauli and correlation hole\textemdash a suppressed probability of finding two fermions near each other. However, as the doping is reduced below a critical value, correlations between such local magnetic moments become positive, signaling strong bunching of doublons and holes. Excellent agreement with numerical linked-cluster expansion (NLCE) and determinantal quantum Monte Carlo (DQMC) calculations is found. Positive non-local moment correlations directly imply potential energy fluctuations due to doublon-hole pairs, which should play an important role for transport in the Fermi-Hubbard model.
\end{abstract}

\maketitle

A central question in understanding cuprate high-temperature superconductors is how spin and charge correlations give rise to the wealth of observed phenomena. Antiferromagnetic order present in the absence of doping quickly gives way to superconductivity upon doping with holes or electrons~\cite{lee06hightc}, suggesting the viewpoint of competing phases. On the other hand, antiferromagnetic correlations can also occur in the form of singlet bonds between neighboring sites, and indeed Anderson proposed~\cite{ande87} that superconductivity could result, upon doping a Mott insulator, from the condensation of such resonating valence bonds. It has also been argued~\cite{lee06hightc} that the pseudo-gap and ``strange metal'' regions are supported by a liquid of spin-singlets. This motivates the simultaneous examination of nearest-neighbor spin and charge correlations, which might reveal the underlying mechanisms of pairing and transport.

In recent years, ultracold atomic gases have been established as pristine quantum simulators of strongly correlated many-body systems~\cite{ingu08varenna,bloc08review,Esslinger2010FermiHubbard}. 
The Fermi-Hubbard model is of special importance due to its paradigmatic role for understanding high-T$_c$ cuprates. At low temperatures and away from half-filling, its theoretical solution presents a severe challenge due to the fermion sign problem. Central properties of Fermi-Hubbard physics, from the reduction of double occupancy~\cite{jord08,taie2012} and compressibility~\cite{schn08,duarte2015} to short-range antiferromagnetic correlations~\cite{Greif2013Magnetism,Hart2015FermiHubbard,Greif2015} and the equation of state~\cite{duarte2015,Cocchi2015,hofrichter2015}, have been observed in ultracold atom experiments.
The recently developed Fermi gas microscopes~\cite{Cheuk2015,Haller2015,Parsons2015,Omran2015,Edge2015,Cocchi2015} have led to the direct observation of 2D fermionic Mott insulators, band insulators, and metals with single-atom, single-site-resolved detection~\cite{Greif2016,CheukMott2016}. The full strength of these microscopes, however, unfolds when single-site detection is used to directly measure correlations in the gas, as achieved with bosons in~\cite{Endres2011stringorder,Endres2013,Islam2015}.
\begin{figure*}[t]
\centering
\includegraphics[scale=0.9]{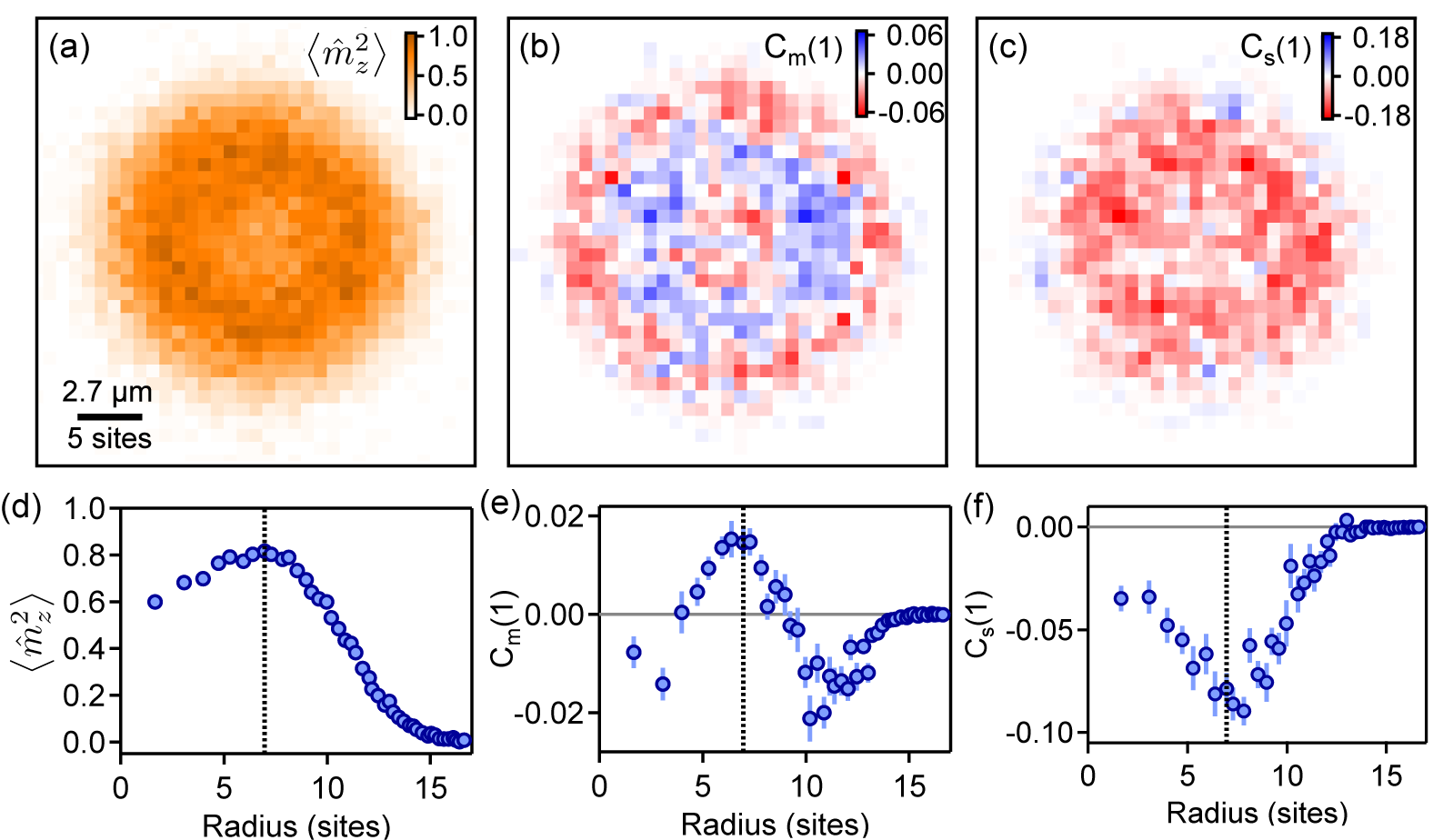}
\caption{Local moment and nearest-neighbor charge and spin correlations in an ultracold atom realization of the Fermi-Hubbard model for $U/t=7.2(1)$. (a,b,c) Averaged local moment, nearest-neighbor moment correlation, and nearest-neighbor spin correlation, respectively, as functions of position. The spatial variations reflect the varying local doping due to the underlying trapping potential. (d,e,f) Radial averages of (a), (b), and (c) respectively. The half-filling point is marked by vertical dotted lines.}
\label{fig1}
\end{figure*}
\begin{figure}[t]
\centering
\includegraphics[scale=1.0]{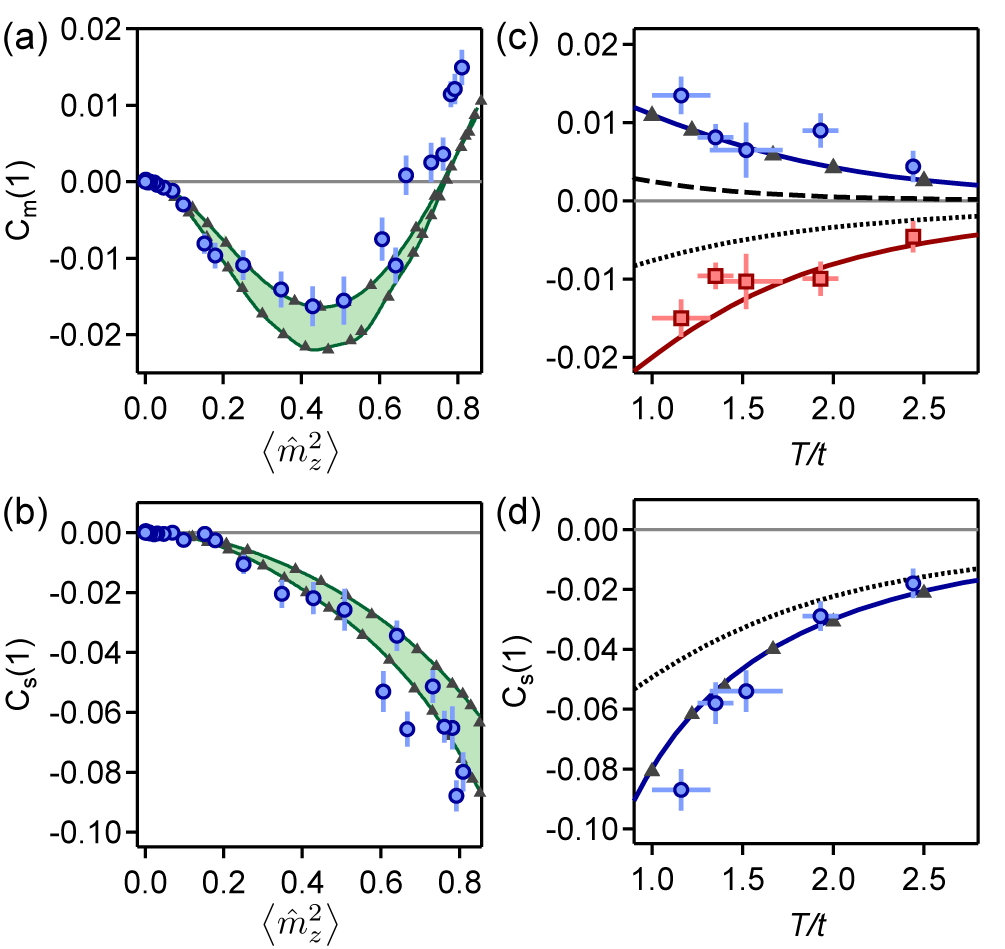}
\caption{Spin and moment correlators as functions of doping and temperature for $U/t=7.2(1)$.  (a,b) Nearest-neighbor moment correlator and spin correlator as functions of the local moment, respectively, shown in blue circles. Results from NLCE (DQMC) for a range of temperatures $T/t=0.89-1.22$ are shown in shaded green (gray triangles). (c) The maximum and minimum of the moment correlator as functions of temperature are shown in blue circles and red squares, respectively. Corresponding results are obtained from NLCE, shown in solid blue and solid red, respectively, for the non-interacting gas, shown in black dashed and dotted lines, respectively, and from DQMC for the correlator at half-filling (gray triangles). (d) Nearest-neighbor spin correlator as a function of temperature (blue circles). Solid blue line: NLCE curve, gray triangles: DQMC; black dotted line: non-interacting gas. For all graphs, theory curves are not adjusted for the experimental imaging fidelity of 95\%.}
\label{fig2}
\end{figure}

In this work, we directly observe charge and spin correlations in the two-dimensional Fermi-Hubbard model using a Fermi gas microscope of $^{40}$K atoms~\cite{Cheuk2015,CheukMott2016}.
Spin correlations displaying antiferromagnetic behavior have also been observed very recently with fermionic $^6$Li in one~\cite{Boll2016} and two~\cite{Parsons2016} dimensions.
We employ the local resolution to simultaneously obtain correlations in the entire range from zero doping (half-filling) to full doping (zero filling), as the density varies in the underlying trapping potential.
The microscope measures the parity-projected density on a given lattice site, i.e. doubly occupied sites (doublons) appear as empty. For a two-spin mixture of fermions in the lowest band of the optical lattice, this is described by the magnetic moment operator~\cite{CheukMott2016} $\hat{m}_{z,i}^2 = (\hat{n}_{\uparrow,i}-\hat{n}_{\downarrow,i})^2$, where $\hat{n}_{\sigma,i} = \hat{c}_{\sigma,i}^\dagger \hat{c}_{\sigma,i}$ is the number operator and $\hat{c}_{\sigma,i}$ ($\hat{c}^{\dagger}_{\sigma,i}$) are fermion annihilation (creation) operators for spin $\sigma =\, \uparrow,\downarrow$ on site $i$. Many repeated images yield the average local moment on each site, see Fig. 1(a,d). The average local moment is a thermodynamic quantity that directly measures the interaction energy of the gas. Indeed, the Fermi-Hubbard Hamiltonian can be written in terms of local moments as
\begin{equation}
  \hat{H} = -t \sum_{\left<i,j\right>,\sigma} \hat{c}_{\sigma,i}^\dagger \hat{c}_{\sigma,j} - \frac{U}{2} \sum_i \hat{m}_{z,i}^2 - \mu \sum_i \left(\hat{n}_{\uparrow,i}+\hat{n}_{\downarrow,i}\right).
\end{equation}
Here, $\left<i,j\right>$ denotes nearest-neighbor sites $i$ and $j$, $t$ is the nearest-neighbor hopping amplitude, $U$ is the on-site interaction energy, and $\mu$ is the chemical potential.
At moderate temperatures and various fillings $n_i=\bra\hat{n}_{\uparrow,i}+\hat{n}_{\downarrow,i}\ket$, this model yields metallic, band insulating, and Mott insulating states. At half-filling ($n_i=1$) and at temperatures below the super-exchange scale $4 t^2/U$, quasi-long-range antiferromagnetic correlations arise. For a fixed temperature, these correlations are expected to be maximal when $U \approx 8 t$, where the interaction energy equals the single-particle bandwidth. Upon doping, a pseudo-gap phase emerges; at even lower temperatures one expects a $d$-wave superconducting state~\cite{lee06hightc}.
While the super-exchange scale is a factor of about two lower than the temperatures achieved here, site-resolved detection of short-range correlations should already reveal precursory signs of physics at this energy scale.

Fig.~1(a) shows a typical measurement of the average local magnetic moment at a given lattice site, from ${\sim} 100$ individual experimental realizations at $U/t{=}7.2 (1)$. Atoms are confined in a radially symmetric trapping potential. Under the local density approximation, this results in a varying local chemical potential, and thus a locally varying filling $n$ throughout the sample. We prepare samples where the maximum filling, which occurs in the center of the trap, lies above $n{=}1$. The half-filling point is then found from radially averaged profiles (see Fig.~1(d)) as the radial position where the moment reaches its maximum. This follows from the particle-hole symmetry of the moment operator $\hat{m}_{z,i}^2$, a property that holds for all its averages and cumulants~\cite{CheukMott2016}.

While fluctuations of the local moment operator do not yield additional information, correlations of the moment on differing sites do~\cite{Kapit2010}. We experimentally measure the moment correlator at a separation of one site, $C_{m}(1)$, defined as
\begin{equation}
C_{m}(1) = \frac{1}{4}\sum_{j \in{\rm nn}_i} \left( \left<\hat{m}_{z,i}^2 \hat{m}_{z,j}^2\right>- \left<\hat{m}_{z,i}^2\right> \left< \hat{m}_{z,j}^2\right>\right),
\end{equation}
where the sum is over all four nearest neighbors.
The locally resolved correlator $C_{m}(1)$ and its radial average are shown in Fig.~1(b,e) respectively.
It displays non-monotonic behavior, changing sign as the filling is lowered. As we discuss below, the negative regions, which indicate anti-bunching of moments, constitute a direct observation of the Pauli and correlation hole. The positive region near maximum moment, i.e. half-filling, reveals instead the bunching of moments, which as shown below arises from an effective attraction between doublons and holes.

Local moment correlations alone, however, are not sensitive to the sign of the spin $\hat{S}_{z,i}{=}\frac{1}{2}\left(\hat{n}_{\uparrow,i}-\hat{n}_{\downarrow,i}\right)$. One important correlator that does depend on the sign of the spin is $\left<\hat{S}_{z,i} \hat{S}_{z,j}\right>$, which can reveal antiferromagnetic ordering, expected to occur at half-filling and at low temperatures. This correlator can be expressed as $\frac{1}{2} \sum_{\sigma}\left<\hat{m}_{\sigma,i} \hat{m}_{\sigma,j}\right> - \frac{1}{4}\left<\hat{m}_{z,i}^2 \hat{m}_{z,j}^2\right>$ (see Supplemental Material), where $\hat{m}_{\sigma,i} = \hat{n}_{\sigma,i} - \hat{n}_{\uparrow,i} \hat{n}_{\downarrow,i}$. All terms can be obtained in separate experimental runs and are separately averaged.
Analogous to the nearest-neighbor moment correlator $C_{m}(1)$, we define the nearest-neighbor spin correlator
\begin{equation}
C_{s}(1) = \sum_{j \in \rm{ nn}_i} \left(\left<\hat{S}_{z,i} \hat{S}_{z,j}\right>- \left<\hat{S}_{z,i}\right>\left<\hat{S}_{z,j}\right>\right).
\end{equation}

Fig.~1(c,f) show the locally resolved nearest-neighbor spin correlation $C_{s}(1)$ and its corresponding radial average. The fact that it is negative suggests antiferromagnetic correlations, as expected~\cite{Hirsch1985,Khatami2011,LeBlanc2013}. However, even without interactions, Pauli-blocking of like spins suppresses $C_s(1)$. 
One can see this by noting that $C_s(1)$ contains density correlations of either spin species separately, $\langle \hat{n}_{\sigma, i} \hat{n}_{\sigma, j} \rangle - \langle \hat{n}_{\sigma, i}\rangle^2$, which are negative even for the non-interacting gas due to Pauli suppression. For the lowest temperatures reached, we observe a maximum absolute spin correlation of about a factor of two larger than that of a non-interacting Fermi gas.

\begin{figure*}[t]
\centering
\includegraphics[scale=1.0]{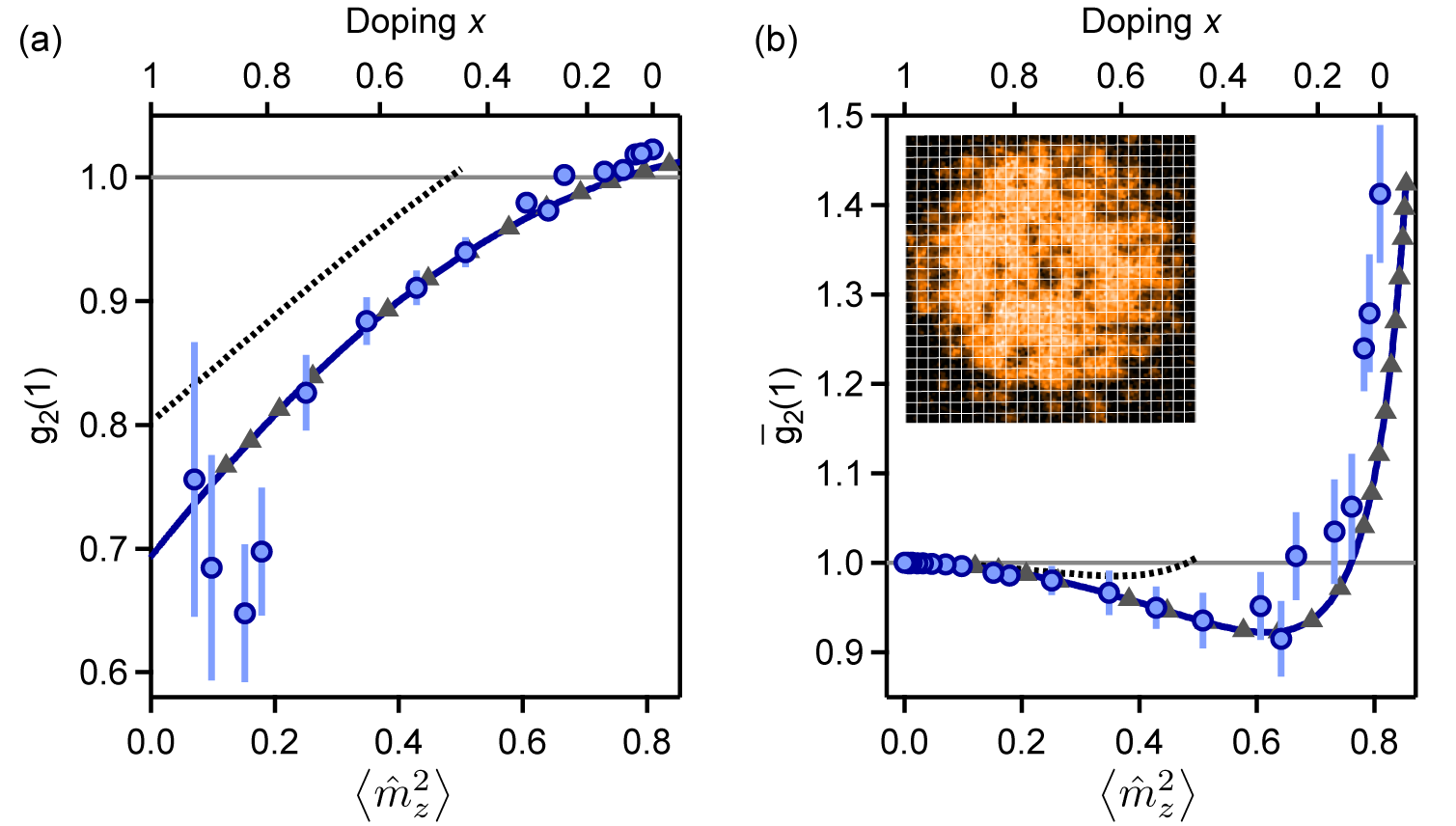}
\caption{Two-point correlation function $g_2$ for moments and $\bar{g}_2$ for anti-moments at a separation of one lattice site for $U/t=7.2(1)$. (a) $g_{2}(1)$ for moments. (b) $\bar{g}_{2}(1)$ for anti-moments. Blue circles: experimental data. Blue solid line: NLCE theory. Gray triangles: DQMC theory. Both NLCE and DQMC calculations are performed at $T/t=1.22$, and are not adjusted for the experimental imaging fidelity of 95\%. Black dotted lines: non-interacting gas. The doping $x$ is zero at maximum moment and one at zero moment; intermediate values of doping as a function of local moment are determined from NLCE theory at $T/t=1.22$, with adjustment for imaging fidelity. Inset: typical image showing neighboring anti-moments (imaged holes) near half-filling. }
\label{fig3}
\end{figure*}

\begin{figure}[t]
\centering
\includegraphics[scale=1.0]{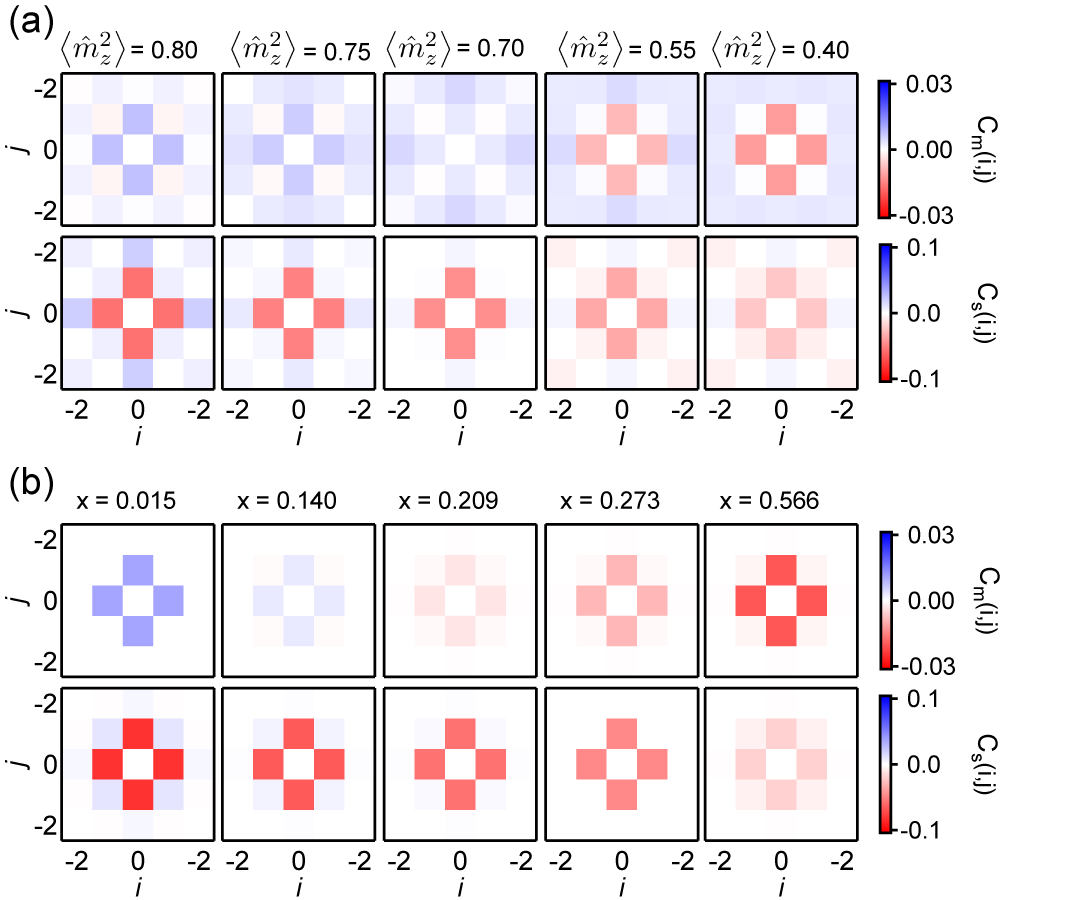}
\caption{Spin and moment correlations as functions of distance and doping. (a) Moment and spin correlations for $U/t=7.2(1)$ are shown in top and bottom row respectively, at various values of the local moment. Correlation values are averaged over symmetric points. The moment correlator $C_m(1,0)$ changes sign near a doping of $x\approx 0.2$.
The anti-correlation of spins $C_s(1,0)$ is observed to weaken upon increasing doping (decreasing moment). In contrast, the next-nearest-neighbor spin correlator $C_s(1,1)$ changes from positive at zero doping to negative at large doping.
(b) Moment and spin correlations obtained from DQMC theory for $U/t=7.2(1)$ and $T/t=1.22$ are shown in top and bottom row respectively, at various values of doping $x$. The nonzero value of the spin correlator at distance $(i,j) = (0,0)$ is omitted for clarity. NLCE results for the spin correlator, not shown, are in agreement with DQMC results.
}
\label{fig4}
\end{figure}

Fig.~2(a,b) show the nearest-neighbor moment and spin correlations versus the measured local moment $\left<\hat{m}_{z,i}^2\right>$.
This representation allows for comparison with theory under minimal assumptions. As a thermodynamic quantity, the moment can replace the role of the chemical potential $\mu$. Indeed all thermodynamic variables can then be viewed as functions of the local moment, the spin correlation at half-filling, $U$, and $t$. In fact, the local spin correlation at half-filling is itself a thermometer that does not require any fit~\cite{ku2012thermodynamics}. Also shown in Fig. 2(a,b) are numerical linked-cluster expansion (NLCE)~\cite{Rigol2006} and determinantal quantum Monte Carlo (DQMC)~\cite{Blanken1981} calculations, which agree with the data within experimental uncertainties. 
Note that there are no free parameters, since the temperature $T/t = 1.16(16)$ is obtained from the spin correlation at half filling.
Concerning the NLCE and DQMC calculations of the moment correlator performed for this work, it is the first time such a high-order correlator (involving terms with up to eight fermion operators) has been calculated with these techniques.

As expected, the antiferromagnetic spin correlations are maximum at half-filling and decrease with increased doping. Moment correlations instead are negative at low to intermediate fillings, crossing zero around a moment of 0.75 (doping $\approx 0.21$) before turning positive towards half-filling. This implies that moments change their character from effectively repulsive (anti-bunching) to effectively attractive (bunching).
The anti-bunching and bunching behaviors in the moments, as well as the antiferromagnetic spin correlations, become more pronounced as the temperature is lowered. Fig.~2(c) shows the positive peak in the moment correlations at half-filling as well as the minimum moment correlation versus temperature; the spin correlator, shown in Fig.~2(d), displays the same behavior, reaching $-0.09$ at our lowest temperatures. This is about 30\% of the maximum spin correlation expected for the spin-1/2 Heisenberg model at zero temperature in 2D~\cite{Paiva2010}.

To interpret the moment correlations, one may recast them in terms of the two-point correlator
\begin{equation}
 g_2(r)=\left<\hat{m}_z^2(r) \hat{m}_z^2(0)\right>/\left<\hat{m}_z^2\right>^2,
\end{equation}
which measures the probability of finding two moments a distance $r$ from each other. In the absence of correlations, $g_2 = 1$. At low filling, where the the doublon density is negligible and the moment $\left<\hat{m}_z^2\right> = \left<\hat{n}\right> - 2 \left<\hat{n}_\uparrow \hat{n}_\downarrow\right>\approx n$ is essentially the density, $g_2(r)$ measures density correlations. These are non-trivial even for the spin-polarized non-interacting Fermi gas, where fermion statistics imply that $g_2(0)=0$, reflecting the fact that two fermions cannot be on the same site. This Pauli suppression of $g_2$ persists at short distances on the order of the average interparticle spacing, a feature known as the Pauli hole. While implications of this fermion ``anti-bunching'' have been observed in the suppression of density fluctuations~\cite{sann10densityfluc,muel10densityfluc} and momentum space correlations~\cite{rom06,Jelt07HBT}, the real space suppression $g_2(r)$ has not been observed in situ before.
In a non-interacting two-spin mixture, the strength of the Pauli hole is halved, as only two identical fermions experience the Pauli hole. Nevertheless, repulsive interactions between opposite spins also suppress $g_2(r)$, leading to a combined Pauli and correlation hole.

In Fig. 3(a), we show the directly measured $g_2(1)$ as a function of moment at an intermediate interaction of $U/t = 7.2$. The strong suppression of $g_2(1)$ at low fillings (large interparticle spacing) is observed, and is stronger than Pauli suppression alone, reflecting short-range anti-correlations due to repulsive interactions.
As shown in Fig.~3(a), the data is well described by NLCE and DQMC calculations.

While $g_2(r)$ describes the probability of finding two moments a distance $r$ from each other, near half-filling, where $\langle \hat{m}_z^2\rangle\sim1$, the correlations arise mainly from sites where the moment is zero, i.e. sites with holes and doublons.  The number of holes and doublons, which appear empty after imaging, is given by $\langle1-\hat{m}_z^2\rangle$. The corresponding two-point correlation function $\bar{g}_2(r)$ of these ``anti-moments'' is thus
\begin{equation}
\bar{g}_2(r)=\left<\left(1-\hat{m}_z^2(r)\right)\left(1- \hat{m}_z^2(0)\right)\right>/\langle1-\hat{m}_z^2\rangle^2.
\end{equation}
In Fig.~3(b), we observe that $\bar{g}_2(1)$ is strongly enhanced near half-filling beyond the uncorrelated value of 1. $\bar{g}_2(1)$ thus reveals the strong bunching of holes and doublons.
There are three contributions to $\bar{g}_2(1)$: correlations between pairs of holes, between pairs of doublons, and between holes and doublons. One expects neighboring holes and neighboring doublons to show negative correlations, due to Pauli suppression and strong repulsion. Hence the bunching behavior must originate from positive correlations between neighboring doublon-hole pairs. This expectation is confirmed by NLCE and DQMC calculations (see Supplemental Material).

The strong doublon-hole correlation near half-filling in the presence of antiferromagnetic correlations can be qualitatively captured by a simple two-site Hubbard model, experimentally realized in~\cite{murmann2015}. While in the strongly interacting limit ($U\gg t$) the doublon density vanishes and the ground state is a spin singlet, at intermediate interaction strengths, tunneling admixes a doublon-hole pair into the ground state wavefunction, with an amplitude ${\sim}t/U$. Thus, short-range singlet correlations at moderate $U/t$ occur naturally together with nearest-neighbor doublon-hole correlations. 

At a separation of one lattice site, we have revealed the competition between Pauli- and interaction-driven repulsion of singly-occupied sites and the effective attraction of doublons and holes, which manifests itself in a sign change of the correlator. The ability of the microscope to measure at a site-resolved level also allows investigation of longer-distance correlations. 
In Fig.~4(a,b), we show the moment and spin correlations $C_{m}(i,j)$ and $C_{s}(i,j)$, respectively, as a function of separation distance $i\hat{x} + j\hat{y}$.
Near half-filling, even at the temperatures of this graph, $T/t \approx 1.2$, antiferromagnetic spin correlations beyond the next neighbor are visible. With increased doping, they give way to a more isotropic negative spin correlation. For example, $C_s(1,1)$ changes sign from positive at half filling to negative at large dopings. This resembles the effect of Pauli suppression present already for non-interacting fermions. For the moment correlator, we observe clearly the sign change of $C_m(1,0)$ at a doping of $x \approx 0.21$, and that the correlations do not extend significantly beyond one site. To our knowledge, these are the first observations and theoretical calculations of spatial moment correlations in the 2D Fermi-Hubbard model.

The measurement of non-local moment correlations also gives direct access to the associated potential energy fluctuations. From the Fermi-Hubbard Hamiltonian in Eq. (1), one finds that
\begin{eqnarray}
  \Delta E_{\rm pot}^2 &=& \frac{1}{4} U^2 \left(\left<\hat{M}^2\right> - \left<\hat{M}\right>^2\right) \nonumber \\
&=& \frac{1}{4} U^2 \sum_{i,j} \left(\left<\hat{m}_{z,i}^2 \hat{m}_{z,j}^2\right>-\left<\hat{m}_{z,i}^2\right>\left<\hat{m}_{z,j}^2\right>\right).
\end{eqnarray}
where $\hat{M} = \sum_i \hat{m}_{z,i}^2$ is the total moment operator.
At half-filling, the contribution to the fluctuations from the nearest-neighbor moment correlations is thus $U^2C_{\rm m}(1) \approx 0.8\, t^2$.
This suggests that doublon-hole correlations can indeed arise from coherent tunneling of particles bound in spin singlets.

The microscopic detection of spatial correlations in both the spin and charge as a function of temperature, doping, and interaction strength in the 2D Fermi-Hubbard model presented here allowed the direct observation of the Pauli and correlation hole at high doping, expected for a Fermi liquid, and of antiferromagnetic correlations accompanied by doublon-hole bunching near half-filling. These correlations were subsequently identified in novel theoretical calculations. It is interesting to note that away from half-filling, both NLCE and DQMC calculations are currently limited to a temperature range around $T/t \approx 0.5$, not far below what is reached experimentally in this work. Further reduction in experimental temperatures will provide a valuable benchmark for theoretical techniques, especially away from half-filling, where the sign problem arises. The clear importance of doublon-hole correlations motivates further studies of their dynamics, especially away from half-filling, which could elucidate their role for the transport properties of a possible ``strange metal'' phase and potential pseudo-gap behavior.

\begin{acknowledgments}
We would like to thank Senthil Todadri, Mohit Randeria, and Markus Greiner and his research group for fruitful discussions.
This work was supported by the NSF, an AFOSR PECASE and MURI on Exotic Quantum Phases, ARO MURI on Atomtronics, and the David and Lucile Packard Foundation. MAN was supported by the DoD through the NDSEG Fellowship Program. KRL was supported by the Fannie and John Hertz Foundation and the NSF GRFP. TP acknowledges support from  CNPQ and FAPERJ. MR was supported by the Office of Naval Research.\\
\\
$^\dagger$ These authors contributed equally to this work.
\end{acknowledgments}

\pagebreak
\clearpage
\setcounter{equation}{0}
\setcounter{figure}{0}
\renewcommand{\thefigure}{S\arabic{figure}}
\onecolumngrid
\begin{center}
\large{\textbf{Supplemental Material: \\ Observation of Spatial Charge and Spin Correlations in the 2D Fermi-Hubbard Model}}
\end{center}
\twocolumngrid

\subsection{Experimental Setup and Procedure}
To prepare fermionic $^{40}$K atoms at low entropies in a square optical lattice, we first sympathetically cool the K atoms in a plugged magnetic trap with bosonic $^{23}$Na. The atoms are subsequently transferred to an optical trap. They are then transported optically into a single 2D layer below our high-resolution microscope objective, where further evaporation occurs~\cite{CheukMott2016}. A square lattice formed by retro-reflected lattice beams, with lattice spacing of $a=541\,\rm{nm}$, is then ramped to a depth that gives the desired Hubbard parameters $U/t$. The Hubbard parameter $U$ and the lattice depth are calibrated with modulation spectroscopy. The tunneling $t$ is extracted from the measured lattice depth using a tight-binding model.

To detect atoms, we freeze the position of the atoms by ramping up the square lattice in $2\,\rm{ms}$ to a depth of $\sim 100 E_R$, where $E_R= \frac{\hbar^2}{2m}\frac{\pi^2}{a^2}$ and $m$ is the mass of $^{40}$K. The square lattice in the $x$-$y$ plane, along with an additional $532\,\rm{nm}$-spaced lattice along $z$, are further ramped up to $\sim 1000 E_R$. We subsequently perform site-resolved fluorescence imaging using Raman sideband cooling. The fluorescence images are then used to reconstruct the parity-projected occupation at each lattice site~\cite{Cheuk2015}.
\begin{figure}[ht]
\centering
\includegraphics[scale=1.2]{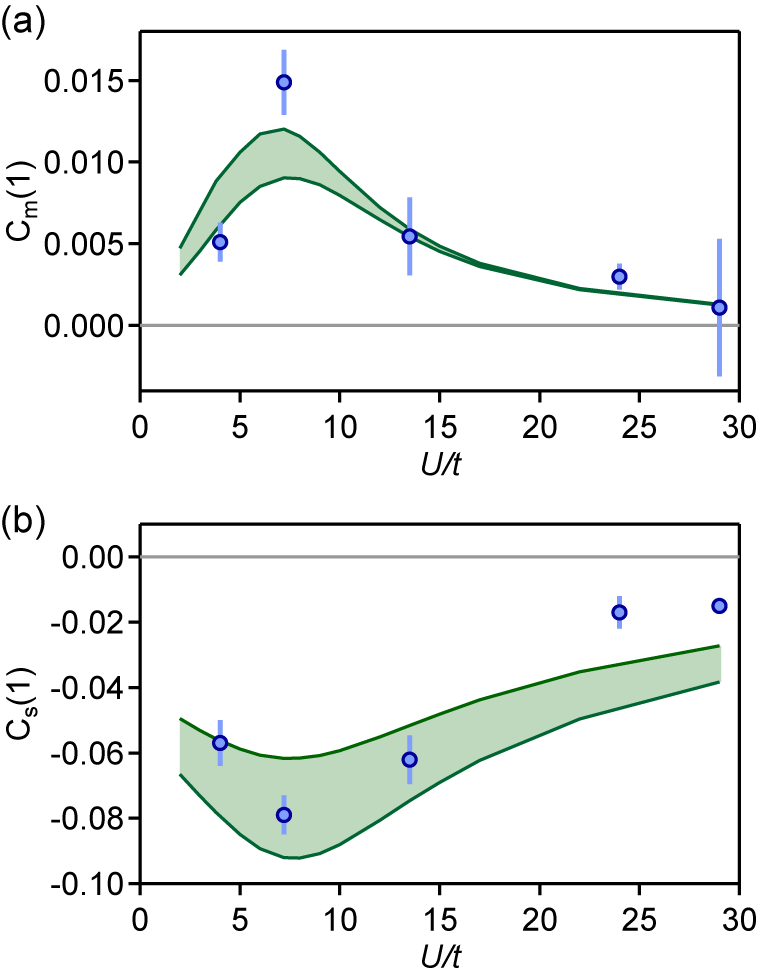}
\caption{Correlations vs $U/t$. (a) Nearest-neighbor moment correlator and (b) nearest-neighbor spin correlator at half-filling are shown in blue circles as functions of $U/t$. Green shaded region indicates NLCE theory curves for the temperature range $T/t = 0.89$ to $1.22$.}
\label{figS1}
\end{figure}

\begin{figure}[t]
\centering
\includegraphics[scale=1.0]{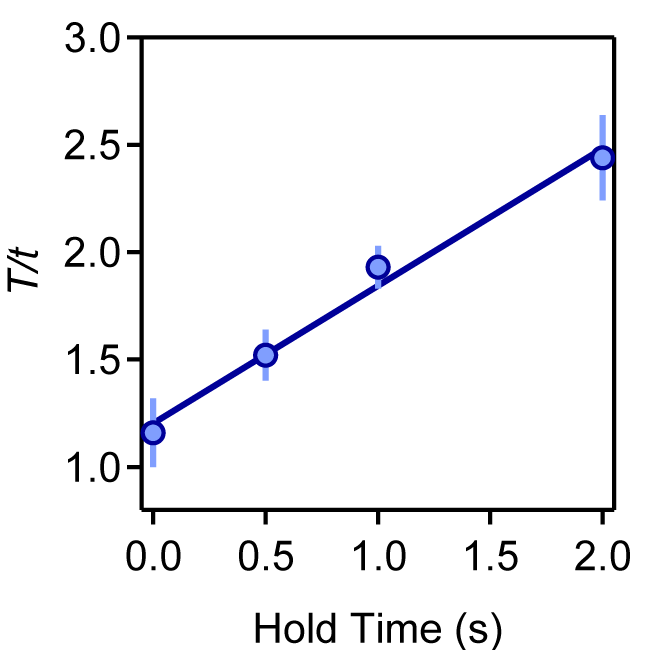}
\caption{Temperature vs hold time in lattice. Blue circles indicate temperature of samples at $U/t = 7.2(1)$ held in the lattice for varying times. Temperature is extracted from the observed moment at half-filling using NLCE theory. A linear fit, shown in blue solid, gives a heating rate of $0.64(5) t/s$.}
\label{figS2}
\end{figure}

\subsection{Spin-sensitive Imaging}
To realize spin-sensitive imaging, we first ramp up the $x$-$y$ lattice to $100\,E_R$ to freeze the distribution. Then we transfer atoms in $\left|F=9/2, m_F=-9/2\ket$ $\left(\left|F=9/2, m_F=-7/2\ket\right)$ to a different hyperfine state $\left|7/2, -7/2\ket$ $\left(\left|7/2, -5/2\ket\right)$ using a microwave sweep. This is followed by a $5\,\rm{ms}$ pulse of light resonant with the $F=9/2\rightarrow F'=11/2$ transition. We measure that this removes $>95\%$ of the atoms in $F=9/2$, while removing only $<0.03\%$ in $F=7/2$. Subsequently the $x$-$y$ lattice and the additional $z$-lattice with $532\,\rm{nm}$ spacing are ramped to $\sim 1000\,E_R$ for Raman imaging. Using samples with a band-insulating core, we have verified that spin-sensitive imaging does not create more than $1(3)\%$ singly occupied sites from doubly occupied sites.

\subsection{Measuring the Spin Correlation Function}
In this section, we show how the spin correlation function $C_s$ can be expressed in terms of experimentally accessible quantities. The spin correlator can be written as
\begin{equation}
C_{s} = 4\bra \hat{S}_{z,i} \hat{S}_{z,j} \ket_c
\end{equation}
where $\bra \hat{A}\hat{B}\ket_c$ denotes the connected part $\bra \hat{A}\hat{B} \ket - \bra \hat{A}\ket \bra \hat{B} \ket$.
We define the following quantities
\begin{eqnarray}
\hat{n}_{i} &=&  \hat{n}_{\uparrow,i}+\hat{n}_{\downarrow,i} \\
\hat{d}_i &=& \hat{n}_{\uparrow,i}\hat{n}_{\downarrow,i} \\
\hat{S}_{z,i} &=& \frac{\left(\hat{n}_{\uparrow,i}-\hat{n}_{\downarrow,i}\right)}{2}
\end{eqnarray}
where $\hat{n}_{\uparrow,i}$ and $\hat{n}_{\downarrow,i}$ denote the on-site spin-up and spin-down number operators, respectively. The directly measured experimental observables are then given by,
\begin{eqnarray}
\hat{m}_{z,i}^{2} &=& \hat{n}_{i}-2\hat{d_i}  \\
\hat{m}_{\sigma,i} &=& \hat{n}_{\sigma,i}-\hat{d_i}
\end{eqnarray}
where $\hat{m}_{z,i}^{2}$ represents the total moment operator on a given lattice site, obtained by imaging both spin components, and $\hat{m}_{\sigma,i}$ represents the observable that we measure via the spin-sensitive imaging technique described above. The correlation signals available are then
\begin{eqnarray}
m^{(2)}\left(i,j\right) &\equiv& \bra\hat{m}_{z,i}^{2} \hat{m}_{z,j}^{2} \ket_c  \\
m_{\sigma}^{(2)}\left(i,j\right) &\equiv& \bra\hat{m}_{\sigma,i} \hat{m}_{\sigma,j} \ket_c
\end{eqnarray}

We begin by showing how the spin-spin correlation function, $4\bra \hat{S}_{z,i} \hat{S}_{z,j} \ket_c$, can be written in terms of these quantities:
\begin{eqnarray}
4\bra \hat{S}_{z,i} \hat{S}_{z,j} \ket_c &=& \bra\left(\hat{n}_{\uparrow,i}-\hat{n}_{\downarrow,i}\right)\left(\hat{n}_{\uparrow,j}-\hat{n}_{\downarrow,j}\right) \ket_c \nonumber \\
&=& \bra\left(\hat{m}_{\uparrow,i}-\hat{m}_{\downarrow,i}\right)\left(\hat{m}_{j\uparrow}-\hat{m}_{\downarrow,j}\right) \ket_c \nonumber \\
 &=& \bra\hat{m}_{\uparrow,i}\hat{m}_{\uparrow,j}\ket_c + \bra\hat{m}_{\downarrow,i}\hat{m}_{\downarrow,j}\ket_c \nonumber \\
& &-\bra\hat{m}_{\uparrow,i}\hat{m}_{\downarrow,j}\ket_c - \bra\hat{m}_{\downarrow,i}\hat{m}_{\uparrow,j}\ket_c.
\label{szsz1}
\end{eqnarray}

Next, we note that $\hat{m}_{z,i}^2 = \hat{m}_{\uparrow,i} +\hat{m}_{\downarrow,i}$, which leads to the following identity:
\begin{eqnarray}
\bra \hat{m}_{z,i}^2 \hat{m}_{z,j}^2 \ket_c &=& \bra\hat{m}_{\uparrow,i}\hat{m}_{\uparrow,j}\ket_c +\bra\hat{m}_{\downarrow,i}\hat{m}_{\downarrow,j}\ket_c \nonumber \\
& &+\bra\hat{m}_{i\uparrow}\hat{m}_{j\downarrow}\ket_c + \bra\hat{m}_{i\downarrow}\hat{m}_{j\uparrow}\ket_c.
\label{szsz2}
\end{eqnarray}
Combining Eq.~(\ref{szsz1}) and Eq.~(\ref{szsz2}) then gives
\begin{eqnarray}
4\bra \hat{S}_{z,i} \hat{S}_{z,j} \ket_c &=& 2\left(\bra\hat{m}_{\uparrow,i}\hat{m}_{\uparrow,j}\ket_c +\bra\hat{m}_{\downarrow,i}\hat{m}_{\downarrow,j}\ket_c\right) \nonumber \\
& &- \bra \hat{m}_{z,i}^2 \hat{m}_{z,j}^2\ket_c
\end{eqnarray}
In terms of the experimentally measured correlators, we thus have
\begin{equation}
C_s = 2(m_{\uparrow}^{(2)}\left(i,j\right)+m_{\downarrow}^{(2)}\left(i,j\right))-m^{(2)}\left(i,j\right)
\end{equation}
Note that this expression is general for any spin imbalance. For the data presented in this work, the ratio of the two spin populations was found to be $1.02(1)$ by comparing the measured average atom number for each spin species after the spin-sensitive imaging.

The above derivation assumes perfect fidelity. There are a few sources of error, one of which is the overall imaging fidelity $f<1$. This reduces the correlation signal by a factor of $f^2$, but does not introduce systematic biases. Neglecting this fidelity $f$, errors that arise from imperfect spin-imaging can be quantified by changing the observable operators from Eqn.~$(6)$ to

\begin{equation}
\hat{\tilde{m}}_{\sigma,i}=\left(1-\epsilon_{1\sigma}\right)\hat{n}_{\sigma,i}+\epsilon_{2\sigma}\hat{n}_{-\sigma,i}-\left(1-\epsilon_{1\sigma}+\epsilon_{2\sigma}\right)\hat{d}_i.
\end{equation}
Here $\epsilon_{1\sigma}$ denotes unintended losses of spin-$\sigma$ atoms when performing spin-sensitive imaging, $\epsilon_{2\sigma}$ denotes imperfect removal of $-\sigma$ atoms, and doubly occupied sites are assumed to be dark. The experimentally available correlators are then given by
\begin{eqnarray}
\tilde{m}^{(2)}\left(i,j\right) &\equiv& \bra\hat{m}_{z,i}^{2} \hat{m}_{z,j}^{2} \ket_c  \\
\tilde{m}_{\sigma}^{(2)}\left(i,j\right) &\equiv& \bra\hat{\tilde{m}}_{\sigma,i} \hat{\tilde{m}}_{\sigma,j} \ket_c
\end{eqnarray}
Neglecting overall imaging fidelity $f$, to leading order in $\epsilon_{1\sigma}$ and $\epsilon_{2\sigma}$, the error in the spin correlator is
\begin{eqnarray}
\Delta C_s &=& \sum_{\sigma} \left\{-(\epsilon_{1\sigma} + \epsilon_{2\sigma} )C_s - (\epsilon_{1\sigma}-\epsilon_{2\sigma})m^{(2)} \right. \nonumber \\
& & \left. - (\epsilon_{1\sigma}-\epsilon_{1-\sigma})(m^{(2)}_{\sigma}-m^{(2)}_{-\sigma})\right\}
\end{eqnarray}
In other words, in the case of perfect non-detection of doubly-occupied sites, each error term is separately proportional to a measured correlator. This allows us to set bounds on the magnitude of these errors. From our measurements, we can bound the first term to $-0.2(1) \times C_s$, the second term to $0.02(2) \times m^{(2)}$, and the third term to $<1\times 10^{-4}$ over all fillings. The latter two terms are negligible compared to the first, as are terms of higher order in $\epsilon_{1\sigma}$ and $\epsilon_{2\sigma}$.

In the case of imperfect non-detection of doubly occupied sites, where we image $\epsilon_d$ of the doublons, to leading order in $\epsilon_d$ one has the additional error term of
\begin{equation}
\epsilon_d \bra 2 \hat{n}_{i} \hat{d}_{j} + 4\hat{d}_i \hat{d}_j\ket_c
\end{equation}
Using an estimate of this correlator from NLCE data, we expect the error to be bounded by 0.025$\epsilon_d$. From the agreement of NLCE data with the correlations versus moment, we estimate that $\epsilon_d<0.10$, giving an error bound of $2.5\times 10^{-3}$.

\begin{figure*}[ht]
\centering
\includegraphics[scale=1.0]{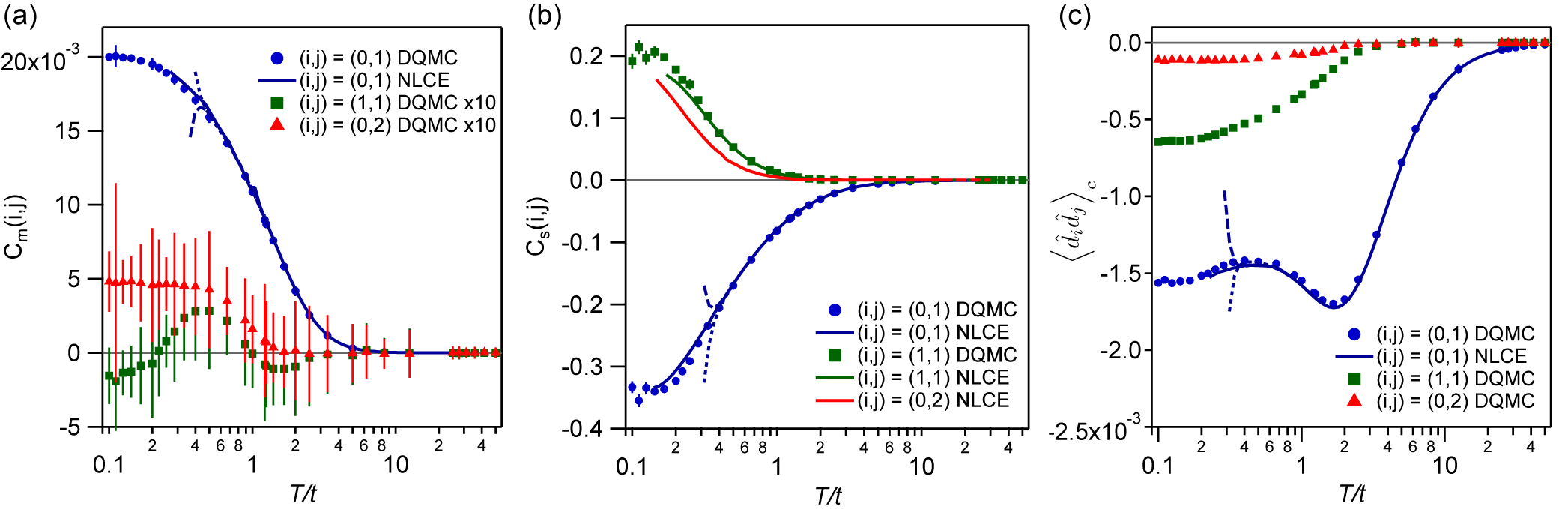}
\caption{Charge and spin correlations versus temperature. Shown are DQMC and NLCE results for (a) the moment correlator $C_m(i,j)$, (b) the spin correlator $C_s(i,j)$, and (c) the doublon-doublon correlator $\langle d_i d_j\rangle_c$ at half-filling as a function of temperature, at $U/t=7.2$. The NLCE results for 8th and 9th orders, and after numerical re-summations, are shown in dotted, dashed and solid lines. Where available, DQMC and NLCE results agree for $T/t>0.5$.}
\label{figS3}
\end{figure*}

\subsection{Data Analysis and Error Estimates}
In each experimental run, the fluorescence image is reconstructed to determine the parity-projected occupation of each lattice site. To obtain the moment $\langle \hat{m}_{z,i}^2\rangle$ at each site, we average over the reconstructed images from $\sim 100$ experimental runs. To obtain the moment correlator $\langle \hat{m}_{z,i}^2 \hat{m}_{z,j}^2 \rangle$, where sites $i$ and $j$ are separated by $(p,q)$ sites, we shift each reconstructed image by $(p,q)$ and multiply the shifted image with the un-shifted image. This multiplied image is then averaged over many experimental runs. A similar procedure is applied to the single-spin images to obtain $\langle \hat{m}_{\sigma,i} \hat{m}_{\sigma,j}\rangle$. For each separation $(p,q)$, the symmetry of the square lattice is employed to provide more statistics. For example, the correlator with $(p,q)=(1,1)$ is averaged with correlators at equivalent separations of $(1,-1)$, $(-1,-1)$ and $(-1,1)$.

Since the trap is radially symmetric, the site-resolved moment $\langle \hat{m}_{z,i}^2\rangle$ can be radially averaged. The radial averaging is performed with a bin size of 40 points. Error bars are statistical, and denote the standard deviation in each bin normalized by the square root of the number of points in each bin. An analogous radial averaging procedure is performed on the correlators $C_s$ and $C_m$, and error bars are similarly obtained. This allows us to replot the correlators as functions of the average moment $\langle \hat{m}_{z,i}^2\rangle$, which is a fit-free method to parametrize the filling.

To obtain the correlators $C_s(i,j)$ and $C_m(i,j)$ at various fillings for Fig. 4 of the main text, the correlators are first plotted versus the average moment. A 6$^{th}$ order polynomial spline is then fit to the data for each separation $(i,j)$. This allows one to interpolate between various fillings.

To determine the temperature of a sample with minimal assumptions, we use the maximum value of the moment as a thermometer. The maximum value of the moment occurs at half-filling, due to particle-hole symmetry. For the temperature range and interaction strength accessed in this work, the moment is monotonic in temperature. Comparison to NLCE theory, accounting for imaging fidelity $f=95(1)\%$, allows us to estimate the temperature of the gas. While this method is not particularly sensitive at low temperatures, it does not rely on assumptions of the trapping potential. We find that using temperatures obtained this way, the behavior of correlators versus temperature agrees with DQMC and NLCE theory.

\subsection{Correlations versus $U/t$}
As shown in Fig.~S1, we observe the largest spin and charge correlations near $U/t\approx 8$, as expected from theoretical predictions. At higher $U/t$, the values of the spin correlation do not lie on an isothermal curve. This could indicate technical heating at the higher lattice depths used, or difficulty reaching thermal equilibrium due to the slower timescale when $t$ is small. To estimate the heating, we hold the gas at $U/t=7.2(1)$ and observe how the temperature changes as a function of hold time. We observe a heating rate of $0.64(5)\,t/\rm{s}$ at this interaction strength, as shown in Fig.~S2.

\subsection{Numerical Linked Cluster Expansion Results}
In numerical linked-cluster expansions~\cite{Rigol2006,Tang2013} one expresses extensive properties of a quantum lattice model in the thermodynamic limit in terms of contributions from finite clusters, up to a certain size, that can be embedded in the lattice:
\begin{equation}
P =  \sum_c M(c) \times W_P(c),
\end{equation}
where $P$ represents the extensive property per site, the sum runs over all clusters that are not related by lattice symmetries, $M(c)$ is the number of ways per site in which each cluster $c$ appears, and
$W_P(c)$ is the contribution of cluster $c$ to property $P$. $W_P(c)$ is calculated using the inclusion-exclusion principle,
\begin{equation}
W_P(c) = p(c) - \sum_{s\subset c} W_P(s),
\end{equation}
where $p(c)$ is the property calculated for cluster $c$ and the sum is over sub-clusters of $c$, all clusters with smaller number of sites that can be embedded in $c$.

NLCEs use the same basis as high-temperature series expansions (HTSEs), however, unlike HTSEs, in which $p(c)$ are expressed as perturbative expansions in terms of the inverse temperature $\beta$, in NLCEs $p(c)$ are calculated exactly (to all orders in $\beta$) using exact diagonalization. Despite the lack of an obvious small parameter in the NLCEs, the convergence of the series is lost below a certain temperature where the correlations in the system grow beyond a length of the order of the largest clusters considered.
Here, we have carried out the site expansion for the Hubbard model ~\cite{Khatami2011,Khatami2012} to the 9th order (maximum cluster size of 9). We work in the grand canonical ensemble and choose a fine chemical potential grid, leading to a high resolution also in the average density~\cite{Rigol2007}. For the correlation functions and temperature ranges of interest in this study, the series shows convergence at all densities; therefore, in the main text, we show results in 9th order only. In Fig. S3, we show the 8th and 9th orders along with results after numerical re-summations using the Wynn algorithm \cite{Tang2013} in an extended temperature region at half filling.

\subsection{Determinantal Quantum Monte Carlo Results}
The determinantal or auxiliary field quantum Monte Carlo (QMC) for interacting fermions on a bipartite lattice is an unbiased algorithm that provides statistically ``exact" answers for the energy and correlation functions as a function of temperature~\cite{Blanken1981, White1989, Santos2003}. By introducing auxiliary fields at each space-time point $\sigma({\bf x},t)$ in a path integral representation of the partition function, the quartic interaction terms can be factored into quadratic forms $\hat{a}_i$. The fermions can now be integrated out yielding a partition function $Z=\sum \rho$ as a sum of determinants $\rho={\rm Tr}_{\{\sigma\}}\prod_{i=1}^{N_\tau} \exp [ \hat{a}_i] $ over auxiliary field configurations \cite{Loh1992,Muramatsu1999}. The negative sign or complex phase of this determinant is the source of the so-called ``sign-problem" in QMC simulations that can limit the ability to go to low temperatures away from half filling. However, in the present case when we compare our QMC with the experimental data for the moment and spin correlators presented in Fig.~S3, we can bracket the lowest temperatures $T$ to be of order the hopping $t$, a regime where the QMC simulations are completely reliable at all fillings. Error bars reported are statistical.

\begin{figure}[h]
\centering
\includegraphics[scale=1.0]{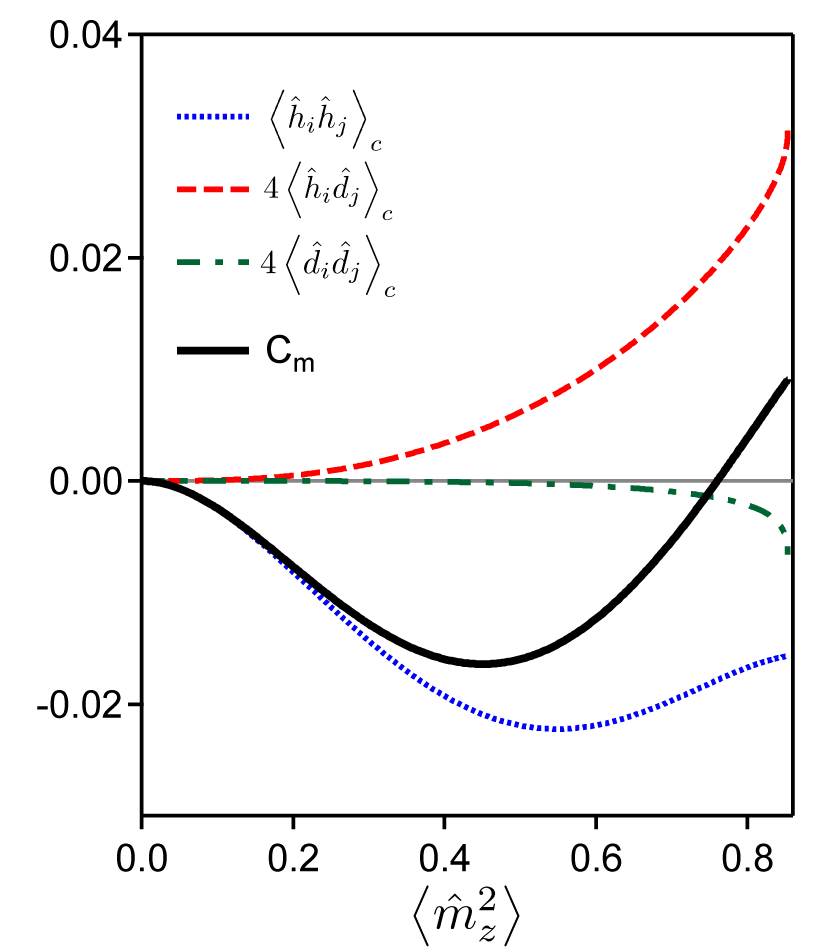}
\caption{Contributions to the nearest-neighbor moment correlator. Shown are the hole-hole (blue dotted), hole-doublon (red dashed), and doublon-doublon (green dot-dashed) contributions, obtained from NLCE theory at $U/t=7.2$ and $T/t=1.22$. The moment correlator, given by the sum of these three terms, is shown in solid black.}
\label{figS4}
\end{figure}

\subsection{Contributions to the Moment-Moment Correlation Function}
Since the local moment $\hat{m}_{z,i}^2 = (\hat{n}_{\uparrow,i} - \hat{n}_{\downarrow,i} )^2$ can be rewritten as $\hat{n}_{i}-2\hat{d}_{i}$, the moment correlator can be expressed in terms of correlators of density and doublons:

\begin{eqnarray}
C_m	&=& \left<\hat{m}_{z,i}^2 \hat{m}_{z,j}^2\right>_c  \nonumber\\
&=&  \left<\hat{n}_{i} \hat{n}_{j}\right>_c- 4  \left<\hat{n}_{i} \hat{d}_{j}\right>_c+  4\left<\hat{d}_{i} \hat{d}_{j}\right>_c.
\label{break}
\end{eqnarray}

Defining the number operator for a spin-$\sigma$ hole at site $i$ to be $\hat{h}_{\sigma,i} = 1- \hat{n}_{\sigma,i}$, and the total hole operator $\hat{h}_i = \hat{h}_{\uparrow,i} + \hat{h}_{\downarrow,i} = 2- \hat{n}_{i}$, one can rewrite the density-doublon correlation $\langle\hat{n}_{i} \hat{d}_{j}\rangle_c$ in terms of holes as $-\langle\hat{h}_{i} \hat{d}_{j}\rangle_c$. Similarly, the density-density correlation can be written as $\langle\hat{n}_{i} \hat{n}_{j}\rangle_c=  \langle\hat{h}_{i} \hat{h}_{j}\rangle_c $. For a homogeneous system, the moment correlator can then be expressed as
\begin{equation}
C_m = \langle \hat{h}_{i} \hat{h}_{j}\rangle_c + 4  \langle \hat{h}_{i} \hat{d}_{j} \rangle_c+  4\langle \hat{d}_{i} \hat{d}_{j}\rangle_c.
\end{equation}
One expects that holes and doublons are repulsive among themselves due to fermion statistics and repulsion, while holes and doublons attract. The moment correlator thus contains the competing effects of inter-hole and inter-doublon repulsion with hole-doublon attraction. The relative weights of the three terms, as extracted from NLCE data, are shown in Fig.~S4 for $U/t=7.2$ and $T/t=1.22$.

\bibliographystyle{apsrev4-1}

\end{document}